\documentclass{pasj01}
\usepackage{graphicx}
\usepackage{natbib}
\usepackage{color}
\Received{$\langle$reception date$\rangle$}
\Accepted{$\langle$acception date$\rangle$}
\Published{$\langle$publication date$\rangle$}

\begin{document}

\title{Hundreds of weak lensing shear-selected clusters from the Hyper Suprime-Cam Subaru Strategic Program S19A data}
\author{Masamune \textsc{Oguri}\altaffilmark{1,2,3}}%
\author{Satoshi \textsc{Miyazaki}\altaffilmark{4,5}}%
\author{Xiangchong \textsc{Li}\altaffilmark{2,3}}%
\author{Wentao \textsc{Luo}\altaffilmark{3}}%
\author{Ikuyuki \textsc{Mitsuishi}\altaffilmark{6}}%
\author{Hironao \textsc{Miyatake}\altaffilmark{7,6,3}}%
\author{Surhud \textsc{More}\altaffilmark{8,3}}%
\author{Atsushi J. \textsc{Nishizawa}\altaffilmark{7}}%
\author{Nobuhiro \textsc{Okabe}\altaffilmark{9,10,11}}%
\author{Naomi \textsc{Ota}\altaffilmark{12}}%
\author{Andr\'{e}s A. \textsc{Plazas Malag\'{o}n}\altaffilmark{13}}%
\author{Yousuke \textsc{Utsumi}\altaffilmark{14}}%
\altaffiltext{1}{Research Center for the Early Universe, University of Tokyo,
  Tokyo, 113-0033, Japan}
\altaffiltext{2}{Department of Physics, University of Tokyo, Tokyo 113-0033, Japan}
\altaffiltext{3}{Kavli Institute for the Physics and Mathematics of the Universe
(Kavli IPMU, WPI), University of Tokyo, Chiba 277-8582, Japan}
\altaffiltext{4}{National Astronomical Observatory of Japan, 2-21-1 Osawa, Mitaka, Tokyo 181-8588, Japan }
\altaffiltext{5}{SOKENDAI (The Graduate University for Advanced Studies), Mitaka,
  Tokyo, 181-8588, Japan}
\altaffiltext{6}{Division of Physics and Astrophysical Science, Graduate School of Science, Nagoya University, Nagoya 464-8602, Japan}
\altaffiltext{7}{Institute for Advanced Research, Nagoya University, Nagoya 464-8601, Japan}
\altaffiltext{8}{The Inter-University Center for Astronomy and Astrophysics (IUCAA), Post Bag 4, Ganeshkhind, Pune 411007, India}
\altaffiltext{9}{Department of Physical Science, Hiroshima University, 1-3-1 Kagamiyama, Higashi-Hiroshima, Hiroshima 739-8526, Japan}
\altaffiltext{10}{Hiroshima Astrophysical Science Center, Hiroshima University, 1-3-1 Kagamiyama, Higashi-Hiroshima, Hiroshima 739-8526, Japan}
\altaffiltext{11}{Core Research for Energetic Universe, Hiroshima University, 1-3-1 Kagamiyama, Higashi-Hiroshima, Hiroshima 739-8526, Japan}
\altaffiltext{12}{Department of Physics, Nara Women's University, Kitauoyanishi-machi, Nara, Nara 630-8506, Japan}
\altaffiltext{13}{Department of Astrophysical Sciences, Princeton University, Peyton Hall, Princeton, NJ 08544, USA}
\altaffiltext{14}{Kavli Institute for Particle Astrophysics and Cosmology, SLAC National Accelerator Laboratory, Stanford University, Menlo Park, CA 94025, USA}
\email{masamune.oguri@ipmu.jp}

\maketitle

\begin{abstract}
We use the Hyper Suprime-Cam Subaru Strategic Program S19A shape
catalog to construct weak lensing shear-selected cluster samples.
From aperture mass maps covering $\sim 510$~deg$^2$ created using a
truncated Gaussian filter, we construct a catalog of 187 shear-selected
clusters that correspond to mass map peaks with the signal-to-noise
ratio larger than 4.7. Most of the shear-selected clusters have
counterparts in optically-selected clusters, from which we estimate
the purity of the catalog to be higher than 95\%. The sample can be
expanded to 418 shear-selected clusters with the same signal-to-noise
ratio cut by optimizing the shape of the filter function and by
combining weak lensing mass maps created with several different
background galaxy selections. We argue that dilution and obscuration
effects of cluster member galaxies can be mitigated by using
background source galaxy samples and adopting the filter function with
its inner boundary larger than about $2'$. The large samples of
shear-selected clusters that are selected without relying on any
baryonic tracer are useful for detailed studies of cluster
astrophysics and cosmology. 
\end{abstract}

\section{Introduction}
Clusters of galaxies are the most massive gravitationally bound
objects in the Universe and have proven to be a key class of objects
for establishing the standard cosmological model that consists of dark
matter and dark energy \citep[see e.g.,][for reviews]{allen11,kravtsov12}.
We can study the internal structure and statistical property of
clusters of galaxies with multi-wavelength datasets, including
optical, X-ray, and radio. For instance, a massive cluster of galaxies
can be securely identified from an overdensity of cluster member
galaxies with similar colors \citep[e.g.,][]{galdders00}. X-ray from
hot gas in clusters of galaxies provides an important means of
finding and studying clusters of galaxies \citep[e.g.,][]{ebeling01}.
Large samples of clusters are being constructed via the
Sunyaev-Zel'dovich effect on cosmic microwave background fluctuations
\citep[e.g.,][]{planck16,hilton21}.

The abundance and internal structure of clusters of galaxies are
mainly determined by the dynamics of dark matter, which makes it
critically important to study the distribution of dark matter in
clusters in great detail. Weak gravitational lensing directly probes
the dark matter distribution in clusters of galaxies and hence plays a
key role in characterizing clusters
\citep[see e.g.,][for a review]{umetsu20}. Among others, weak
gravitational lensing plays an essential role for the use of the
cluster population as a probe of cosmological parameters, because
cluster observables must be linked to cluster masses in order to
compare the observed abundance of clusters with theoretical
predictions \citep[see e.g.,][for a review]{pratt19}. Indeed, attempts
to use clusters as an accurate cosmological probe have often been
hampered by the uncertainty of mass calibrations of clusters for which
complicated selection biases in cluster surveys must be taken into
account.

A new approach that has been explored less extensively is finding
clusters directly from weak gravitational lensing shear data by
identifying peaks in weak lensing mass maps
\citep[e.g.,][]{schneider96,white02,hamana04,hennawi05,maturi05,maturi10,fan10,marian12,lin16}. 
The abundance of the peaks contains information on the abundance of
massive dark matter halos and hence on cosmological parameters
\citep[e.g.,][]{jain00,dietrich10,maturi10,shan14,liu15a,liu15b,hamana15,kacprzak16,shan18}.
In addition, the relatively simple and clean selection function of
weak lensing shear-selected clusters \citep[e.g.,][]{hamana12,chen20}
enables to use them for better understanding of cluster astrophysics.
For instance, the X-ray analysis of shear-selected clusters suggests
that they tend to be X-ray underluminous compared with clusters found
in other techniques \citep{giles15,miyazaki18b}.

However a challenge lies in the requirement of wide and deep imaging
for finding a significant number of weak lensing shear-selected
clusters. First attempts identified only a handful of such clusters,
if restricted to those with a sufficiently high signal-to-noise ratio
(see e.g., Appendix~\ref{app:ti_filter}) of $\gtrsim 5$ 
\citep{wittman01,miyazaki02,hetterscheidt05,wittman06,miyazaki07,schirmer07,gavazzi07,utsumi14}.
On the other hand, more systematic search of weak lensing shear-selected
clusters is made possible thanks to recent progress of wide-field
imaging surveys. For instance, \citet{shan12} constructed a sample of 51
shear-selected clusters with signal-to-noise ratios larger than 4.5 from
the Canada-France-Hawaii Telescope Legacy Survey \citep{heymans12}.

Hyper Suprime-Cam \citep[HSC;][]{miyazaki18a}, which is a wide-field
optical imager mounted on the Subaru 8.2-meter telescope, offers a
unique opportunity for constructing a large sample of weak lensing
shear-selected clusters \citep{miyazaki15}. In particular, the HSC 
Subaru Strategic Program
\citep[HSC-SSP;][]{aihara18a,aihara18b,aihara19}, which is a deep
multi-band imaging survey of 1400~deg$^2$ of the sky, is an ideal
survey for this purpose. \citet{miyazaki18b} presented a sample of 65
weak lensing shear-selected clusters with signal-to-noise ratios
larger than 4.7 from the HSC-SSP first year shape catalog covering
$\sim 160$~deg$^2$ \citep{mandelbaum18a}. Using the same HSC S16A data,
\citet{hamana20} constructed a sample of 124 shear-selected clusters 
with signal-to-noise ratios larger than 5 by mitigating the dilution
effect of foreground and cluster member galaxies.

In this paper, we present updated catalogs of weak lensing
shear-selected clusters from the latest HSC-SSP S19A shape catalog
covering  $\sim 433$~deg$^2$ (Li et al., in prep.). We construct
catalogs using three different approaches adopting different shapes of
filters. We also assign redshifts of individual clusters by cross
matching the shear-selected clusters with optically-selected clusters.

This paper is organized as follows. The data used for our analysis are
summarized in section~\ref{sec:data}. Our method to construct mass
maps is detailed in section~\ref{sec:map}. We present our main results
in section~\ref{sec:results}, and conclude in
section~\ref{sec:conclusion}. Throughout the paper we assume the
matter density $\Omega_{\rm m}=0.3$, the cosmological constant
$\Omega_\Lambda=0.7$, the baryon density $\Omega_{\rm b}=0.05$, the
dimensionless Hubble constant $h=0.7$, the spectral index
$n_{\rm s}=0.96$, and the normalization of the matter power spectrum
$\sigma_8=0.81$. 

\section{Data}\label{sec:data}

\subsection{Weak lensing shape catalog}\label{sec:shape_catalog}

The HSC-SSP S19A shape catalog (Li et al., in prep.) is constructed in
a manner similar to the HSC-SSP S16A shape catalog presented in
\citet{mandelbaum18a} that takes a moment-based approach of
\citet{hirata03}. The multiplicative and additive biases are derived
with realistic image simulations \citep{mandelbaum18b}. The detailed
systematics tests presented in Li et al., in prep. indicate that the
S19A shape catalog is sufficiently accurate and is ready for
various cosmological and astrophysical analyses. The catalog contains
$\sim 36$~million galaxies.

Photometric redshifts of individual galaxies are also measured
with various methods using the HSC $grizy$-band photometry
\citep[see][for photometric redshifts of galaxies in the S16A
  data]{tanaka18}. In this paper we may use subsamples of the S19A
shape catalog that are defined based on these photometric redshifts. 
Specifically, we define subsamples using the so-called P-cut method
\citep[see][]{oguri14,medezinski18}, for which galaxies satisfying 
\begin{equation}
  \int_{z_{\rm min}}^{z_{\rm max}}P(z) dz > P_{\rm th},
  \label{eq:p-cut}
\end{equation}
where $P(z)$ the probability distribution function (PDF) of the
photometric redshift of each galaxy, are included. 
This P-cut allows us to securely and flexibly select
galaxies behind clusters of interest by choosing the parameters
$z_{\rm min}$, $z_{\rm max}$, and $P_{\rm th}$ appropriately
\citep[see][]{medezinski18}, and hence mitigate the dilution effect by
cluster member galaxies \citep[see also][]{hamana20}.
The performance of the P-cut method is shown to be comparable to the
selection of background galaxies in color-color space \citep{medezinski18}.

Throughout the paper we adopt the {\tt dNNz} photometric redshift
measurement (Nishizawa et al., in prep.) for defining background
galaxy samples with the P-cut method. The {\tt dNNz} is based on the
multiple-layer perceptron that consists of six hidden layers where
each layer contains 100 nodes. Input attributes are {\tt cmodel}
magnitude, size, and point spread function matched aperture magnitude
in five broad bands, leading to 15 attributes in total, for each
galaxy. The outputs are probabilities of the galaxy lying at the
redshift bin spanning from $z=0$ to $7$ divided into 100  bins. The
code is trained to minimize the total difference between the output
probabilities and input delta-function like probability summed over
all training sample and redshift ranges. From the analysis on the test
sample that has not been used for training, it is found that redshifts
are accurate at the bias of $10^{-4}$, the scatter of $3\%$, and the
outlier fraction of less than $10\%$ . The {\tt dNNz} can avoid the
overfitting even when the part of the data is missing (e.g., due to
the lack of images in some broadband filters), by introducing
a dropout layer right after the input layer with missing rate being
0.2. When applying the {\tt dNNz} on the real data, the {\tt dNNz}
with the dropout layer is applied only when the part of the data is
missing, otherwise the {\tt dNNz} without the dropout layer is
employed for the complete data to maximize the performance of the
photometric redshift measurement. 

\subsection{Optically-selected cluster catalogs}\label{sec:cluster_catalog}

We use several optically-selected cluster catalogs to assign redshifts
for individual peaks in weak lensing mass maps. First, we adopt a
cluster catalog (S20A version 1\footnote{While the S20A version 2
CAMIRA cluster catalog for which photometry corrections to mitigate
various photometry issues \citep{aihara19} are included is currently
available, in this paper we use the version 1 catalog that was the
latest version when the shear-selected cluster catalogs were
constructed.}) constructed with the photometric data
from the HSC-SSP S20A internal data release using the CAMIRA algorithm
\citep{oguri14}. In \citet{oguri18a}, the CAMIRA algorithm has been
applied to the HSC-SSP S16A data covering $\sim 230$~deg$^2$ to
construct a catalog of 1921 clusters at redshift $0.1<z<1.1$ and
richness $N$ greater than 15 \citep[see][for the definition of
  richness $N$ in the CAMIRA algorithm]{oguri14}.
In this paper, we adopt an updated catalog of 8910 clusters at
redshift $0.1<z<1.38$ and  richness greater than
15 from the HSC-SSP S20A data covering $\sim 830$~deg$^2$.

Since the HSC-SSP survey region is chosen to overlap with the Sloan
Digital Sky Survey \citep[SDSS;][]{york00}, we also use cluster
catalogs from the SDSS for assigning cluster redshifts. 
Specifically, we use the redMaPPer cluster catalog \citep{rykoff14}
that contains clusters at $0.08<z<0.6$ as well as the WHL15 cluster
catalog \citep{wen12,wen15} that contains clusters at $0.05<z<0.79$.
Both of these cluster catalogs were constructed based on the photometric
galaxy catalog covering $\sim 14,000$~deg$^2$ from SDSS Data Release 8
\citep[DR8;][]{aihara11}. In addition to these purely
optically-selected clusters, we also adopt the CODEX cluster catalog
\citep{finoguenov20} that contains clusters at $0.05<z<0.69$ from
ROSAT all-sky survey \citep[RASS;][]{voges99} with optical confirmations
using the redMaPPer algorithm applied to the SDSS DR8 data. Throughout
the paper we use X-ray centroids as centers of the CODEX clusters.

All the cluster redshifts we adopt throughout this paper are
photometric redshifts of clusters derived from the HSC $grizy$-band
photometry for CAMIRA or from the SDSS $ugriz$-band photometry for
redMaPPer, WHL15, and CODEX. The typical accuracy of these cluster
photometric redshifts is $\sigma_z/(1+z)\sim 0.01$, which is
sufficiently accurate for our current purpose. 

\section{Mass maps}\label{sec:map}

\subsection{Introduction}

Since both convergence $\kappa$ and shear $\gamma$ are derived by the
second derivatives of the lens potential, we can derive a convergence
(mass) map from the shear map by a convolution of the shear map with a
kernel \citep{kaiser93}. More generally, in the flat sky coordinate
$\boldsymbol{\theta}$, we can construct a map of the aperture mass
$M_{\rm ap}(\boldsymbol{\theta})$ \citep{schneider96}, which is a
convergence convolved with a spatial filter $U$  
\begin{equation}
M_{\rm ap}(\boldsymbol{\theta}) 
=\int d\boldsymbol{\theta}' \kappa(\boldsymbol{\theta}')
U(|\boldsymbol{\theta}-\boldsymbol{\theta}'|).
\label{eq:m_ap_1}
\end{equation}
Provided that the spatial filter is compensated
\begin{equation}
  \int d\theta\,\theta\,U(\theta)=0,
  \label{eq:compensate}
\end{equation}
it is equivalent to the convolution of the tangential shear with a kernel $Q$
\begin{equation}
  M_{\rm ap}(\boldsymbol{\theta})=\int d\boldsymbol{\theta}'
  \gamma_+(\boldsymbol{\theta'};\boldsymbol{\theta})Q(|\boldsymbol{\theta}-\boldsymbol{\theta'}|),
\label{eq:m_ap_2}
\end{equation}
where $\gamma_+(\boldsymbol{\theta'};\boldsymbol{\theta})$ is the
tangential shear at $\boldsymbol{\theta'}$ defined with respect to
$\boldsymbol{\theta}$ and $Q$ is related with $U$ as
\begin{equation}
Q(\theta)=\frac{2}{\theta^2}\int_0^\theta d\theta'\theta' U(\theta')-U(\theta).
  \label{eq:u2q}
\end{equation}
In this paper, we
consider two types of the spatial filters. One is a truncated Gaussian
filter, which resembles the one adopted in \citet{miyazaki18b} to
construct a shear-selected cluster sample from the HSC S16A data
\citep[see also][]{hamana20}. The other is a filter introduced by
\citet{schneider96}, which we call a truncated isothermal filter
throughout the paper and is designed to optimize the detection of
halos from mass maps. In the next subsections, we describe these
filters in more detail.

\subsection{Truncated Gaussian filter}\label{sec:filter_tg}

We use the following kernel function to define the truncated Gaussian
filter 
\begin{equation}
  Q(\theta)=\frac{1}{\pi\theta^2}
  \left[1-\left(1+\frac{\theta^2}{\theta_0^2}\right)e^{-\theta^2/\theta_0^2}\right]
  e^{-\theta^4/\theta_{\rm out}^4},
  \label{eq:q_tg}
\end{equation}
which is smoothly truncated at $\theta=\theta_{\rm out}$. Thus the
filter is slightly different from the one adopted in
\citet{miyazaki18b} and \citet{hamana20} for which the kernel function
is sharply truncated at $\theta=\theta_{\rm out}$. We adopt this
smoothly truncated form for the numerical stability of our approach to
derive mass maps using the Fast Fourier Transform (see
subsection~\ref{sec:procedure}). The corresponding spatial filter
$U(\theta)$ can be derived as
\begin{equation}
U(\theta)=-Q(\theta)-\int_0^\theta d\theta'\frac{2}{\theta'}Q(\theta').
\end{equation}
Throughout the paper we adopt
$\theta_0=1\farcm 5$ and $\theta_{\rm out}=13'$ so that resulting
signal-to-noise ratios roughly matches those in \citet{miyazaki18b}
and in \citet{chen20}. Following \citet{miyazaki18b}, we also do not
apply any cut in the source galaxy sample. Hereafter this set-up is
referred to as TG15. 

\subsection{Truncated isothermal filter}\label{sec:filter_ti}

\citet{schneider96} introduced the following form of the spatial
filter for the aperture mass
\begin{equation}
  U(\theta)=
  \left\{
  \begin{array}{ll}
    1 & (\theta \leq \nu_1\theta_R)\\
    \frac{1}{1-c}\left(
    \frac{\nu_1\theta_R}{\sqrt{(\theta-\nu_1\theta_R)^2+(\nu_1\theta_R)^2}}-c\right)
    & (\nu_1\theta_R \leq \theta \leq \nu_2\theta_R)\\
    \frac{b}{\theta_R^3}(\theta_R-\theta)^2(\theta-\alpha\theta_R)
    & (\nu_2\theta_R \leq \theta \leq \theta_R)\\
    0 & (\theta_R \leq \theta)
\end{array}
\right.
\end{equation}
where $c$, $b$, $\alpha$ are determined from the condition that
$U(\theta)$ and its first derivative are continuous at
$\theta=\nu_2\theta_R$ as well as the compensation condition given by
equation~(\ref{eq:compensate}). This means that the shape of
$U(\theta)$ is specified by three parameters, $\nu_1$, $\nu_2$, and
$\theta_R$. The corresponding kernel function is derived using
equation~(\ref{eq:u2q}). 

This filter has several desirable properties. First, we can tweak the
shape of the filter quite flexibly by adjusting the three parameters
$\nu_1$, $\nu_2$, and $\theta_R$. Second, it has $Q(\theta)=0$ at
$\theta<\nu_1\theta_R$ and hence allows us to efficiently remove the
contribution from the innermost part of halos, which is a source of
various systematic effects such as the dilution effect by cluster
member galaxies, the effect of reduced shear, and the magnification
bias, to the signal. Third, the filter is confined within a finite
radius (i.e., $U(\theta)=Q(\theta)=0$ at $\theta>\theta_R$) and hence
mitigate the impact of e.g., the boundary of the survey region to the
map.

\begin{table*}[t]
  \caption{Summary of set-ups to construct shear-selected cluster
    samples. }
  \begin{center}
    \begin{tabular}{ lccccc } \hline\hline
      Name & Filter & Parameter values & Source galaxy selection & Num. of galaxies\\ \hline
      TG15 & Truncated Gaussian & $\theta_0=1\farcm5$, $\theta_{\rm out}=13'$ & No cut & $35804886$ \\
           \hline 
      TI05 & Truncated isothermal & $\nu_1=0.021$, $\nu_2=0.36$, $\theta_R=23\farcm8$ & $z_{\rm min}=0.2$, $z_{\rm max}=7$, $P_{\rm th}=0.95$ & 32750421 \\
           & Truncated isothermal & $\nu_1=0.025$, $\nu_2=0.36$, $\theta_R=20\farcm0$ & $z_{\rm min}=0.3$, $z_{\rm max}=7$, $P_{\rm th}=0.95$ & 26024748 \\
           & Truncated isothermal & $\nu_1=0.027$, $\nu_2=0.36$, $\theta_R=18\farcm5$ & $z_{\rm min}=0.5$, $z_{\rm max}=7$, $P_{\rm th}=0.95$ & 20664165 \\
           & Truncated isothermal & $\nu_1=0.027$, $\nu_2=0.36$, $\theta_R=18\farcm5$ & $z_{\rm min}=0.7$, $z_{\rm max}=7$, $P_{\rm th}=0.95$ & 15564223 \\
           \hline
      TI20 & Truncated isothermal & $\nu_1=0.095$, $\nu_2=0.36$, $\theta_R=21\farcm1$ & $z_{\rm min}=0.2$, $z_{\rm max}=7$, $P_{\rm th}=0.95$ & 32750421 \\
           & Truncated isothermal & $\nu_1=0.110$, $\nu_2=0.36$, $\theta_R=18\farcm2$ & $z_{\rm min}=0.3$, $z_{\rm max}=7$, $P_{\rm th}=0.95$ & 26024748 \\
           & Truncated isothermal & $\nu_1=0.121$, $\nu_2=0.36$, $\theta_R=16\farcm6$ & $z_{\rm min}=0.5$, $z_{\rm max}=7$, $P_{\rm th}=0.95$ & 20664165 \\
           & Truncated isothermal & $\nu_1=0.121$, $\nu_2=0.36$, $\theta_R=16\farcm6$ & $z_{\rm min}=0.7$, $z_{\rm max}=7$, $P_{\rm th}=0.95$ & 15564223 \\
           \hline
 \end{tabular}
\end{center}
\label{tab:filters}
\end{table*}

\begin{figure}
  \begin{center}
      \includegraphics[width=0.95\hsize]{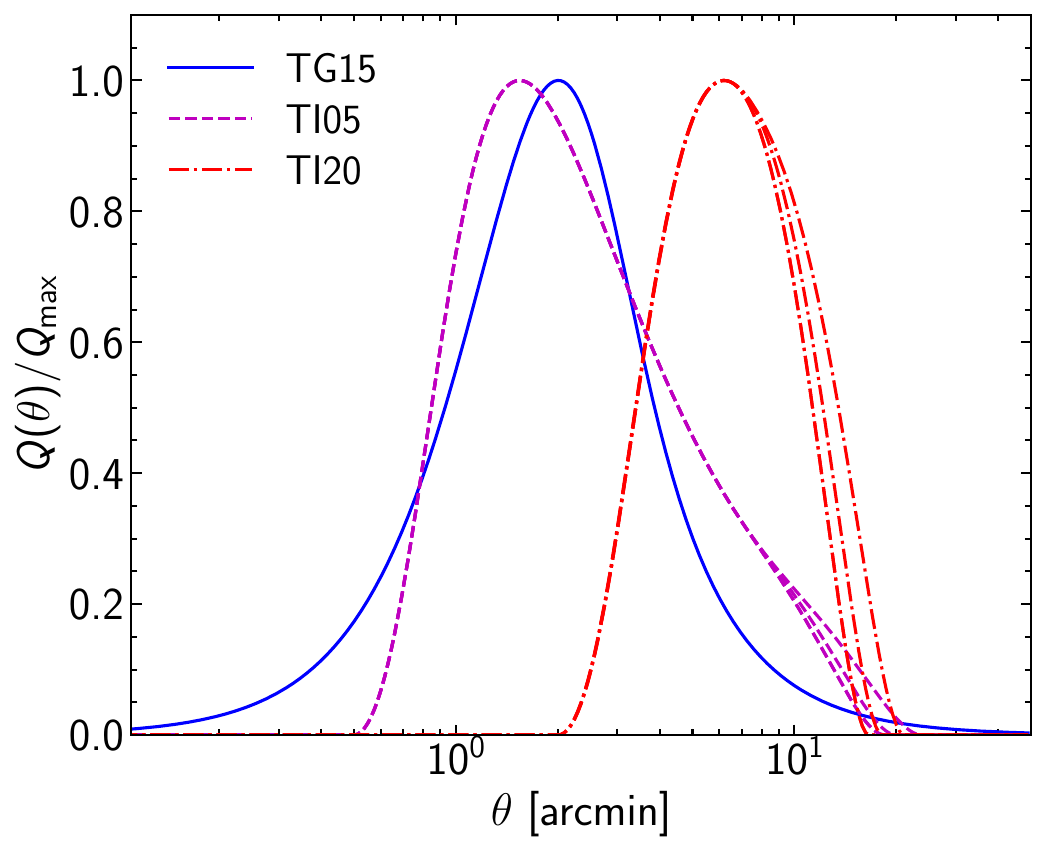}
  \end{center}
  \caption{
    The kernel function $Q(\theta)$, which is used to derive the
    aperture mass map (see equation~\ref{eq:m_ap_2}), is shown. Solid,
    dashed, and dash-dotted lines are $Q(\theta)$ for the TG15, TI05,
    and TI20 set-ups summarized in Table~\ref{tab:filters}. Note that
    we adopt slightly different shapes of $Q(\theta)$ for different
    source galaxy selections. We show $Q(\theta)$ normalized by its
    maximum value.
   }
\label{fig:filter}
\end{figure}

To explore the impact of the different inner boundary of the filter,
$\nu_1\theta_R$, in this paper we consider two different values of the
inner boundary, $\nu_1\theta_R=0\farcm5$, and $2'$, which are referred
as TI05 and TI20, respectively. For both TI05 and TI20, we construct
mass maps  with four different source galaxy subsamples defined
using photometric redshifts of source galaxies (see
subsection~\ref{sec:shape_catalog}), in order to enhance the
detection efficiency particularly at high redshifts \citep[see
  also][]{hamana20}. For each inner boundary of the filter and the
source galaxy sample, we carefully choose parameter values of the
filter to maximize the expected signal-to-noise ratio and to mitigate
the impact of density fluctuations along the line-of-sight on cluster
finding. The specific procedure of the optimization of the parameters
is detailed in Appendix~\ref{app:ti_filter}. Table~\ref{tab:filters}
summarizes set-ups to construct shear-selected cluster samples.
The shapes of the kernel function $Q(\theta)$ used in this paper are
also presented in Figure~\ref{fig:filter}.

\subsection{Practical procedure}\label{sec:procedure}

Following \citet{oguri18b}, we adopt the Fast Fourier Transform (FFT)
to derive aperture mass maps from the shape catalog. For a given
source galaxy sample, we first create a shear map in a two-dimensional
rectangular grid by a simple tangent plane projection. Throughout the
paper, a pixel scale of the grid ($\boldsymbol{\theta}$-coordinate)
of $\Delta\theta=0\farcm25$ is adopted. We use a discrete version of
equation~(\ref{eq:m_ap_2}) with weights to derive an aperture mass map
i.e.,   
\begin{equation}
  M_{\rm ap}(\boldsymbol{\theta}_i)=
  \frac{\left(\Delta\theta\right)^2}{W(\boldsymbol{\theta}_i)}\sum_j 
  w_j\left\{\gamma_+(\boldsymbol{\theta}_j;\boldsymbol{\theta}_i)-c_{+,j}\right\}
  Q(|\boldsymbol{\theta}_i-\boldsymbol{\theta}_j|),
  \label{eq:m_ap_3}
\end{equation}
\begin{equation}
  W(\boldsymbol{\theta}_i)
  =\frac{\sum_j w_j(1+m_j)Q(|\boldsymbol{\theta}_i-\boldsymbol{\theta}_j|)}{\sum_j Q(|\boldsymbol{\theta}_i-\boldsymbol{\theta}_j|)},
\end{equation}
where $i$ and $j$ label each grid, $w_i$ is the weight of $i$-th grid
that is computed from the sum of weights of sources galaxies in the
grid, and $c_{+,i}$ and $m_i$ denote the additive and multiplicative
biases in the $i$-th grid, respectively. We evaluate these summations
by FFT with an appropriate zero padding beyond the boundary. We also
note that, due to the tangent plane projection, the principal axes of
the $\boldsymbol{\theta}$-coordinate are not necessarily aligned with
the North and West directions, even for the HSC-SSP S19A patches (see
below) whose areas are relatively small. 
Since the shear in the HSC-SSP shape catalog is defined with respect
to the equatorial coordinate system, we rotate the shear so that the
shear is converted to the one defined with respect to the
$\boldsymbol{\theta}$-coordinate.  Furthermore, since the
operation to create an aperture mass map (equation~\ref{eq:m_ap_3})
is confined within a relatively small sky area thanks to the compact
size of the convolution kernel $Q(\theta)$, the flat-sky approximation
is expected to be locally accurate in deriving each pixel value of the
aperture mass map.

Since the HSC-SSP S19A shape catalog consists of six disjoint patches
(XMM, VVDS, WIDE12H, GAMA09H, GAMA15H, and HECTOMAP), we create mass
maps for each of these patches to search for peaks.

We define signal-to-noise ratios of peaks using local estimates of
the shape noise from the ``sigma map'' \citep{oguri18b}. We derive the
sigma map by randomly rotating the orientations of source galaxies before
constructing the mass map, and repeating this procedure 500 times. The
sigma map is given by the square root of the variance of the
randomized mass maps. The signal-to-noise ratio $\nu$ is defined by
the ratio of the peak value of the mass map to the noise value at the
peak position from the sigma map. We note that the sigma map derived
by this procedure includes only the shape noise and hence does not
include cosmic shear from the large-scale structure (see also
Appendix~\ref{app:ti_filter}).

We mask the boundary of the survey region as follows. We first derive
the smoothed number density map by first deriving the pixelized number
density maps and then smoothing it by a Gaussian kernel with a
standard deviation of $8'$. The average number density is derived from
the smoothed map with $3\sigma$ clipping. We mask pixels that have
values less than 0.5 times the average number density. We also mask
pixels with values of the sigma map more than 1.5 times higher than
the average value.  

From the map of the signal-to-noise ratio $\nu$, we select peaks with
$\nu\geq 4.7$, which is the threshold adopted also in
\citet{miyazaki18b}. To avoid double counting of clusters, we discard
any peaks that have other peaks with higher $\nu$ within $4'$ from the
peaks.  For mass maps with the truncated isothermal filter (TI05 and
TI20), we create mass maps with four different values of $z_{\rm
  min}$ as shown in Table~\ref{tab:filters}. We create a list of peaks
for each value of $z_{\rm min}$, and combine four lists of peaks with
matching radius of $4'$. For a peak that is detected in multiple mass
maps from different source galaxy selections, we adopt the highest
value of $\nu$ among the mass maps as the signal-to-noise ratio of
that peak.

\begin{figure*}
  \begin{center}
      \includegraphics[width=0.28\hsize]{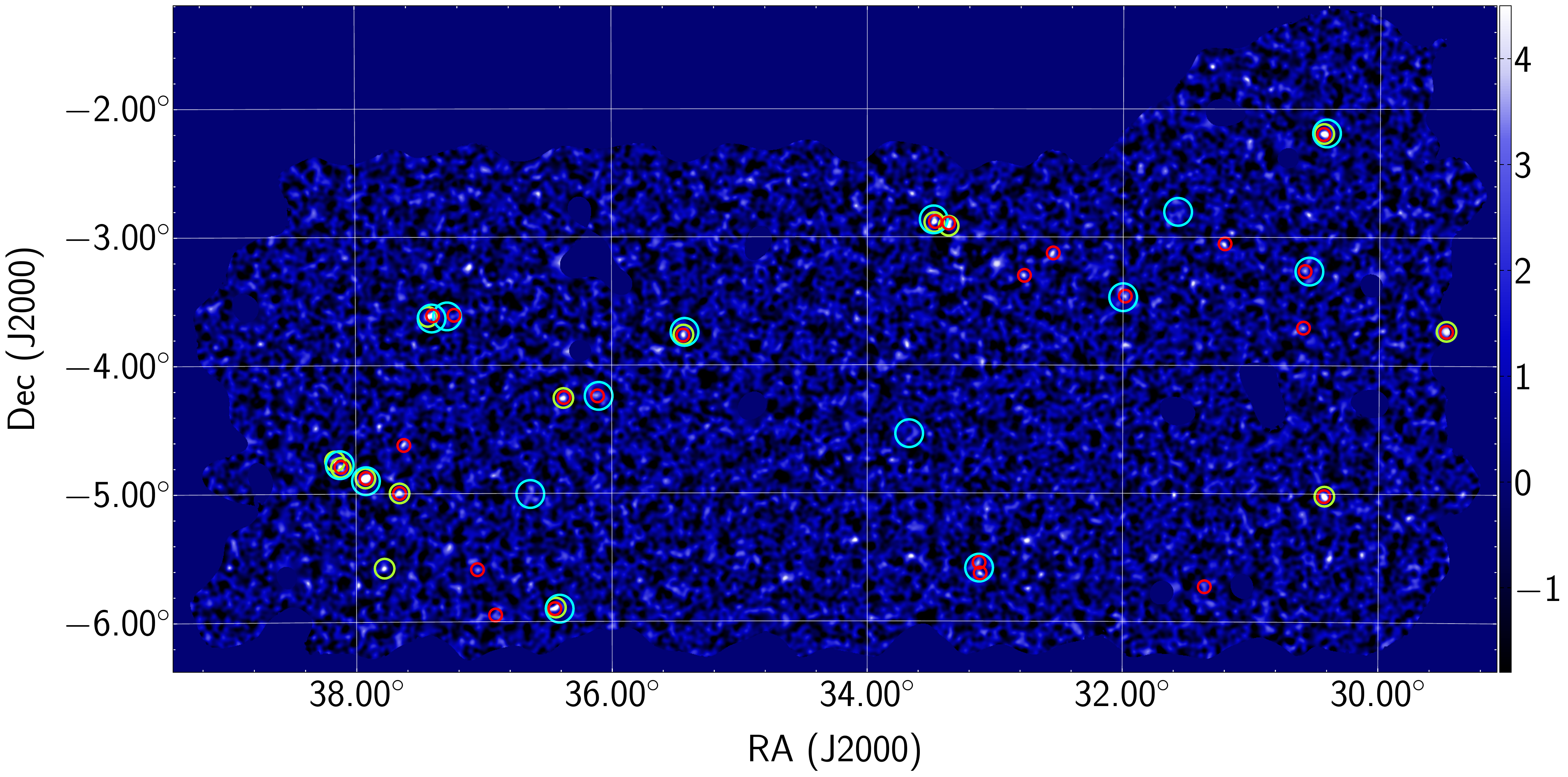}
      \includegraphics[width=0.70\hsize]{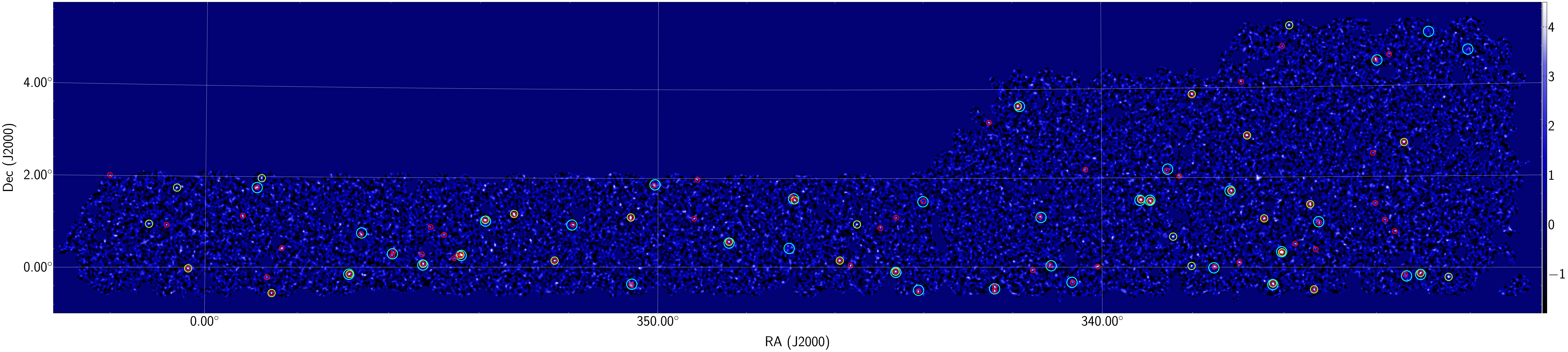}
      \includegraphics[width=0.98\hsize]{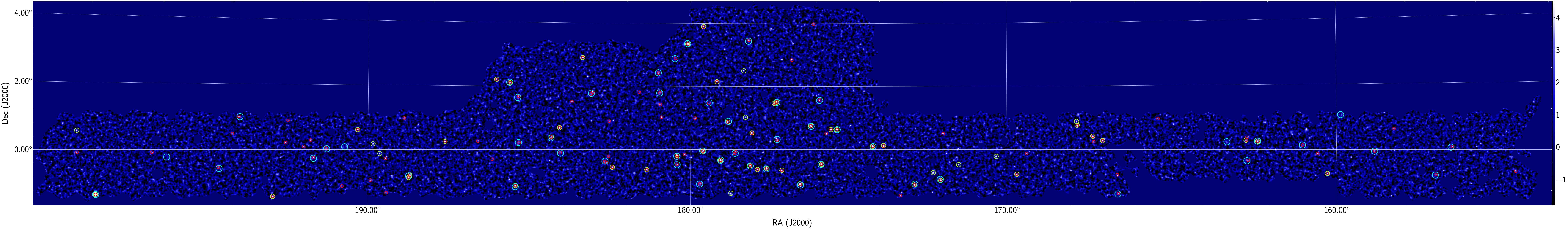}
      \includegraphics[width=0.49\hsize]{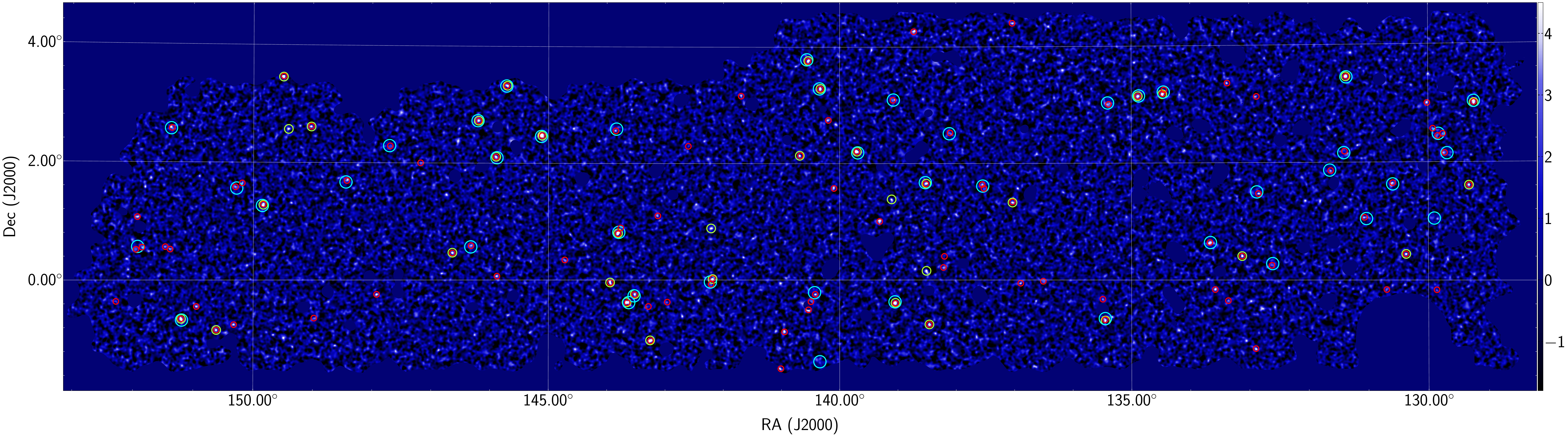}
      \includegraphics[width=0.49\hsize]{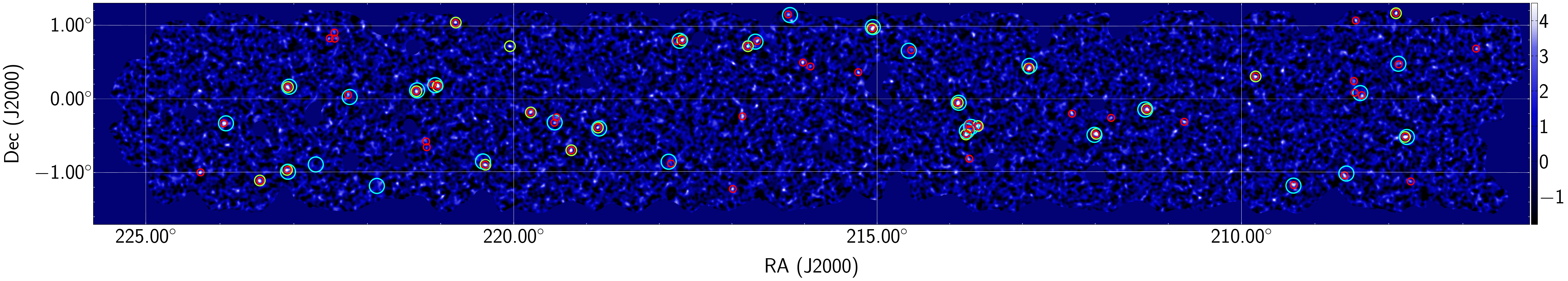}
      \includegraphics[width=0.72\hsize]{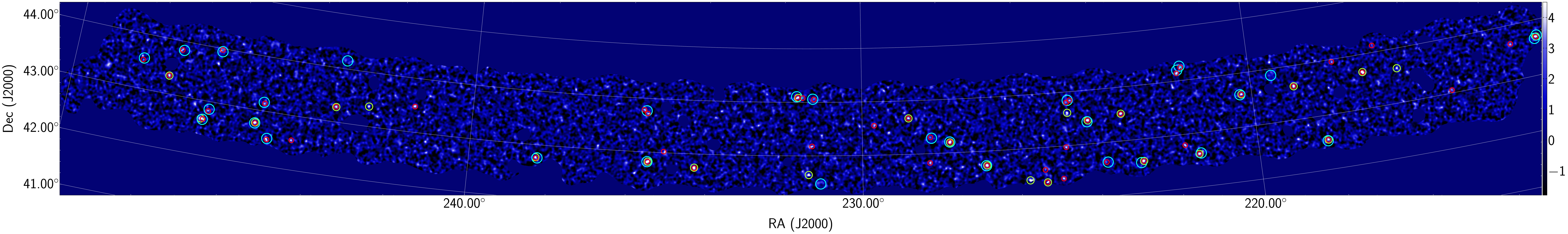}
  \end{center}
  \caption{The spatial distribution of identified shear-selected
    clusters i.e., peaks in mass maps, in six disjoint patches.
    Middle green, small red, large cyan circles show clusters by TG15,
    TI05, and TI20 set-ups (see Table~\ref{tab:filters}),
    respectively. They are overlaid in mass maps created by the
    Gaussian filter without the truncation
    ($\theta_{\rm out}\rightarrow \infty$ in equation~\ref{eq:q_tg})
    with $\theta_0=2'$.
   }
\label{fig:map}
\end{figure*}

\begin{table*}[t]
  \caption{Summary of shear-selected cluster samples and matching with
    optically-selected cluster catalogs. The last column shows the
    number of shear-selected clusters that are matched with any of
    optically-selected cluster catalogs and thus with redshift
    assignments. } 
  \begin{center}
    \begin{tabular}{ lcccccc } \hline\hline
      Name & Num. of clusters & CAMIRA match$^{*}$ & redMaPPer match$^{*}$ & WHL15 match$^{*}$ & CODEX match$^{*}$ & Any match \\ \hline
TG15 & 187 & 163 (21) & 135 (7) & 177 (50) & 59 (6) & 182 \\
TI05 & 418 & 353 (51) & 271 (11) & 364 (104) & 124 (6) & 392 \\
TI20 & 200 & 173 (33) & 138 (10) & 184 (69) & 79 (3) & 193 \\
\hline
    \end{tabular}
\end{center}
\tabnote{ $^{*}$ The number in parentheses indicates shear-selected
  clusters that are matched with multiple optically-selected clusters.}
\label{tab:catalogs}
\end{table*}

\section{Results}\label{sec:results}

\subsection{Shear-selected cluster catalogs and redshift assignments}

We construct shear-selected cluster catalogs from peaks with
$\nu\geq 4.7$ in mass maps covering $\sim 510$~deg$^2$ following the
procedure described in subsection~\ref{sec:procedure}. The catalogs
are constructed for the three set-ups (TG15, TI05, and TI20)
summarized in Table~\ref{tab:filters}. The catalogs contain 187, 418,
and 200 clusters for TG15, TI05, and TI20, respectively. Compared with
the TG15 set-up adopted in \citet{miyazaki18b}, we find roughly twice
the number of shear-selected clusters for TI05, because the shape of
the kernel function for TI05 follows the expected tangential shear
profile more closely than for TG15 and therefore is more optimal.
The small number of shear-selected clusters for TI20, on the other
hand, is due to the removal of the large central region ($<2'$) where
a significant tangential shear signal is observed for many clusters.
Figure~\ref{fig:map} shows the spatial distribution of these
clusters from peaks in mass maps. We note that, even though the numbers
of shear-selected clusters are similar between TG15 and TI20, roughly
half of the clusters are detected both in TG15 and TI20, partly because
most of the clusters have signal-to-noise ratios near the threshold
(see below).

To check whether they are indeed associated with concentrations of red
galaxies and to assign redshifts to these clusters, we cross match
the shear-selected cluster catalogs with four optically-selected
cluster catalogs, HSC CAMIRA, SDSS redMaPPer, SDSS WHL15, and SDSS
CODEX (see subsection~\ref{sec:cluster_catalog} for concise
descriptions of these catalogs). We regard any optically-selected
clusters that are located within the physical transverse distance of
1$h^{-1}$Mpc computed at the cluster redshifts from each
shear-selected cluster as matched clusters. It is possible that
multiple optically-selected clusters are matched with a single
shear-selected cluster, and in that case we regard an
optically-selected cluster that is located closest to the
shear-selected cluster, whose location refers to a peak position in a
mass map, in terms of the physical transverse distance as a primary
match and assign the photometric redshift of the primarily matched
cluster to the shear-selected cluster.    

The result of matching is summarized in Table~\ref{tab:catalogs}.
For TG15 and TI20, $\sim 97$\% of shear-selected clusters have
counterparts in optically-selected cluster catalogs and hence have
redshift assignments. The fraction slightly decreases to $\sim 94$\%
for TI05. For comparison, we generate a random catalog of
shear-selected clusters by randomly drawing points in unmasked
regions of mass maps with the number density of 5~deg$^{-2}$,
and apply the same method for matching shear-selected clusters with
the optically-selected clusters to find that $\sim 33\%$ of
the random shear-selected clusters are matched with any of
optically-selected clusters. Since this fraction represents the chance
probability of matching with the optically-selected clusters, 
we can argue that the true fraction of unmatched shear-selected
clusters may be as high as $3\times 3/2\sim 5$\% for TG15 and TI20,
and $6\times 3/2\sim 9$\% for TI05. Taking also the incompleteness of
our optically-selected cluster samples used for matching (e.g., no
cluster at $z<0.05$) into consideration, we can  argue that the purity
of our shear-selected cluster catalogs is higher than 95\% for TG15
and TI20 and more than 91\% for TI05.  
The catalogs of the shear-selected clusters for TG15, TI05, and TI20
set-ups including the results of cross matching are shown in
Supplementary Tables 1, 2, and 3. 

\begin{figure}
  \begin{center}
      \includegraphics[width=0.95\hsize]{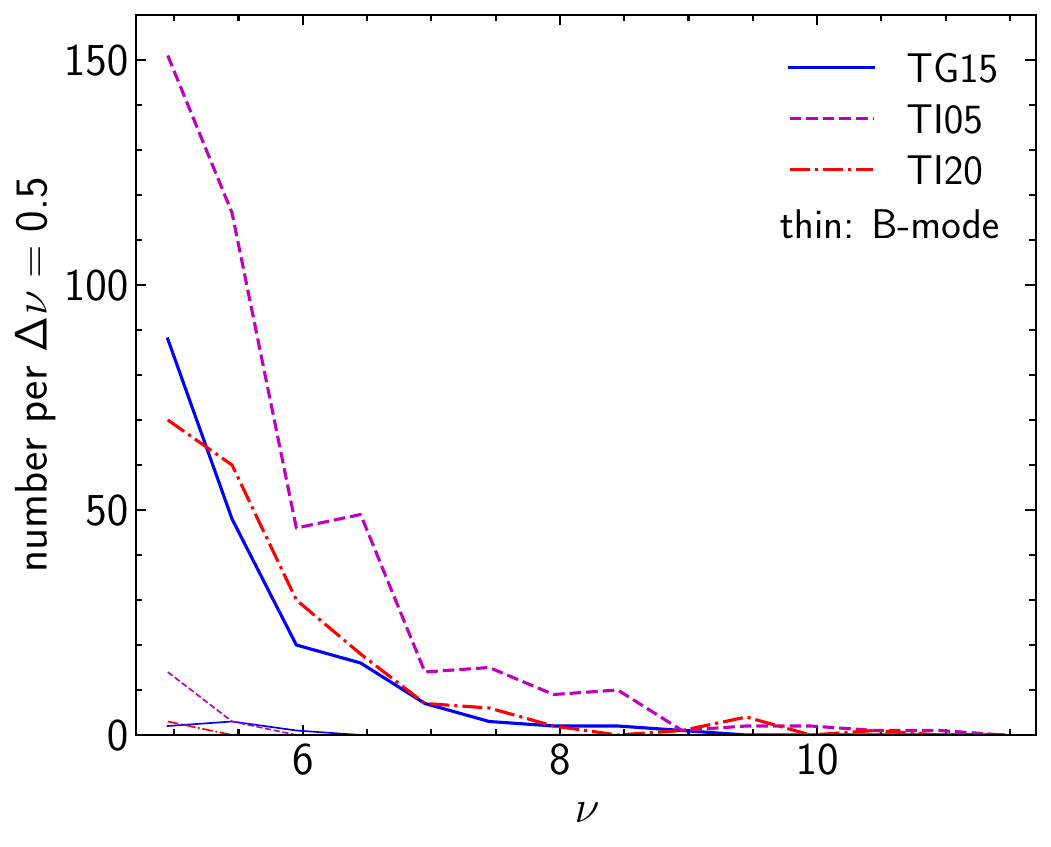}
  \end{center}
  \caption{Distributions of the signal-to-noise ratio $\nu$. Solid,
    dashed, and dash-dotted lines show distributions for the TG15,
    TI05, TI20 set-ups summarized in Table~\ref{tab:filters},
    respectively. Thin lines indicates distributions of the
    signal-to-noise ratio $\nu$ from B-mode mass map peaks.
   }
\label{fig:hist_sn}
\end{figure}

\begin{figure}
  \begin{center}
      \includegraphics[width=0.95\hsize]{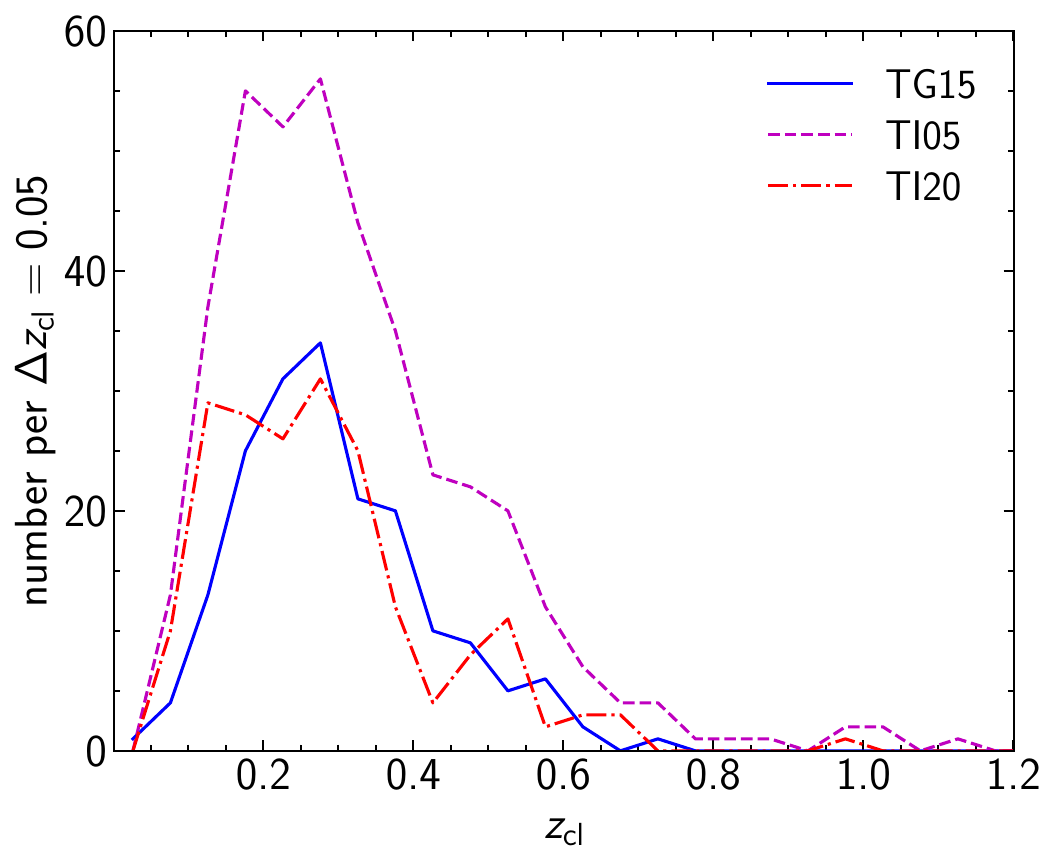}
  \end{center}
  \caption{Similar to Figure~\ref{fig:hist_sn}, but distributions of
    cluster redshift $z_{\rm cl}$ are shown. 
   }
\label{fig:hist_z}
\end{figure}

\begin{figure}
  \begin{center}
      \includegraphics[width=0.95\hsize]{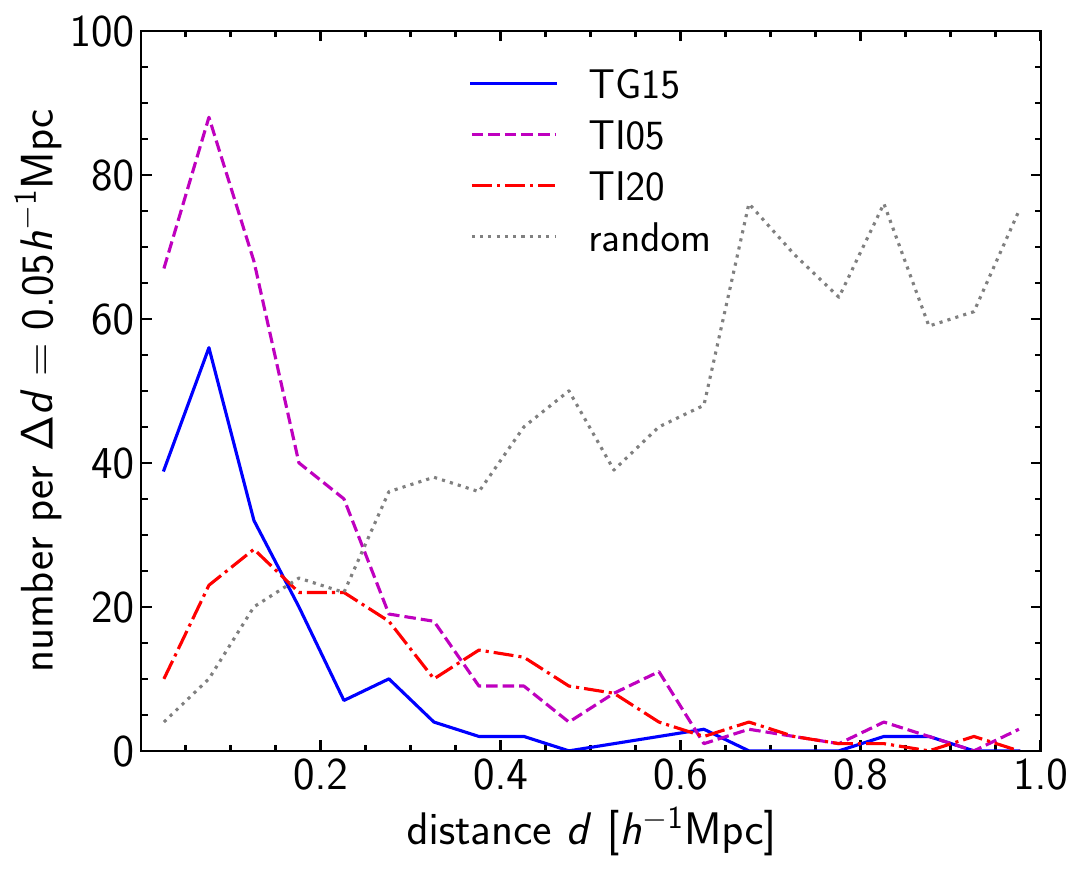}
  \end{center}
  \caption{Similar to Figure~\ref{fig:hist_sn}, but distributions of 
    the physical transverse distance $d$ between the shear-selected
    cluster and the primarily matched optically-selected cluster are
    shown. The distribution of $d$ for the case using the random
    catalog is also shown by the dotted line for reference.
   }
\label{fig:hist_dis}
\end{figure}

We show distributions of the signal-to-noise ratio $\nu$ and the
redshift $z_{\rm cl}$ of shear-selected clusters for all the three
set-ups in Figures~\ref{fig:hist_sn} and \ref{fig:hist_z},
respectively. Number counts rapidly decrease with increasing $\nu$ and
the redshift distributions are peaked at $z_{\rm cl}\sim 0.2-0.3$,
both of which are consistent with theoretical predictions
\citep[see e.g.,][]{miyazaki18b}.

While weak lensing mass maps are constructed from E-mode shear, B-mode
mass maps generated from B-mode shear provide an important means of
checking the validity of the analysis \citep[e.g.,][]{utsumi14}.
As a sanity check, we select mass map peaks from B-mode mass maps
adopting the same signal-to-noise ratio threshold of $\nu\geq 4.7$.
The distribution of B-mode mass map peaks are also shown in
Figures~\ref{fig:hist_sn}. In total there are 6, 17, and 3 B-mode mass
map peaks with $\nu\geq 4.7$ for TG15, TI05, and TI20, respectively.
We find that the numbers of B-mode mass map peaks are sufficiently
small, $\lesssim 4$\%, compared with those of E-mode mass map peaks,
which support the high purity of our shear-selected cluster catalogs
as estimated above. 

In Figure~\ref{fig:hist_dis}, we show distributions of the physical
transverse distance between the shear-selected cluster and the
primarily matched optically-selected cluster. We find that in most
case the distance is small, $\lesssim 0.3h^{-1}$Mpc. The mean distance
is higher for TI20 than in the other set-ups, which can be understood
by the conservative choice of $Q(\theta)$ to remove the small-scale
information (see Figure~\ref{fig:filter}), which naturally leads to
the degraded angular resolution of the resulting mass maps. For
comparison, we also plot the distribution in the case of using the
random catalog mentioned above. Since the spatial distribution of the
random catalog is not correlated with any of optically selected
clusters, the resulting distribution indicates that expected for
matching by chance.  We find that the distribution for the random
catalog differs considerably from those for shear-selected clusters,
supporting the high purity of our shear-selected cluster catalogs.

\begin{figure}
  \begin{center}
      \includegraphics[width=0.95\hsize]{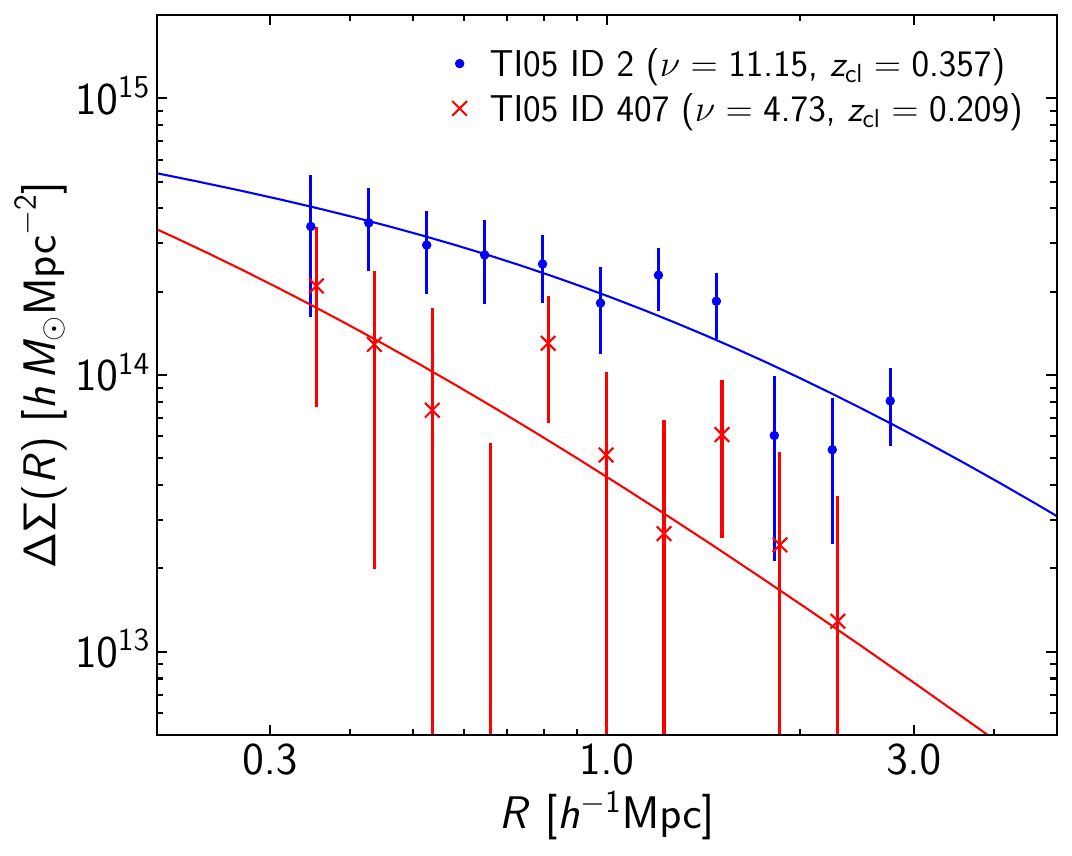}
  \end{center}
  \caption{Examples of tangential shear profiles and fitting them with
    an NFW profile. We show examples for high and low signal-to-noise
    ratio $\nu$ from the TI05 catalog. Symbols with errors show
    observed tangential shear profiles, and lines indicate best-fit
    NFW profiles. Errors include only the shape noise.
   }
\label{fig:wlprofile}
\end{figure}

\subsection{Weak lensing mass measurements}

Following \citet{miyazaki18b}, we derive weak lensing masses of all the
shear-selected clusters with redshift assignments by fitting their
differential surface density profiles. We use the P-cut method
\citep{oguri14,medezinski18} to securely select background galaxies
for each shear-selected cluster, adopting $z_{\rm min}=z_{\rm cl}+0.2$ 
and $P_{\rm th}=0.98$ in equation~(\ref{eq:p-cut}). We derive the
differential surface density profile $\Delta\Sigma(R)$ by fully taking
account of the PDF of the photometric redshift of each galaxy
\citep[see e.g.,][for a specific procedure]{medezinski18}, again
adopting the {\tt dnnz} photometric redshift measurements.
For TG15, we derive differential surface density profiles in the
range $R=[0.3,\,7]h^{-1}$Mpc with a spacing of $\Delta\log R =0.09$.
The outer radius is chosen to be same as that used in
\citet{miyazaki18b} and \citet{chen20} so that the mass bias derived
in \citet{chen20} can be applied. We adopt a slightly larger inner
boundary of $R=0.3h^{-1}$Mpc to mitigate the dilution effect by
cluster member galaxies \citep{medezinski18}. For TI05 and TI20, we
adopt a more conservative radius range of $R=[0.3,\,3]h^{-1}$Mpc. In
all cases, we consider the shape noise and ignore the cosmic shear
error, again to follow the set-up assumed in \citet{chen20}.

The differential surface density profiles are fitted with an NFW
profile \citep{navarro97}, which describes differential surface
density profiles of clusters in numerical simulations reasonably well
out to $\sim 10$ times the virial radius \citep{oguri11}. We
parameterize the NFW profile by $M_{\rm 500c}$ and $c_{\rm 500c}$,
which describe the mass and concentration parameter for the critical
overdensity of $500$. We restrict the range of the concentration
parameter to $0.5<c_{\rm 500c}<10$ and derive both $M_{\rm 500c}$ and
$c_{\rm 500c}$ from fitting to the observed differential surface
density profile of each cluster. We show some examples of our
tangential shear profile fitting in Figure~\ref{fig:wlprofile}.

\begin{figure*}
  \begin{center}
      \includegraphics[width=0.32\hsize]{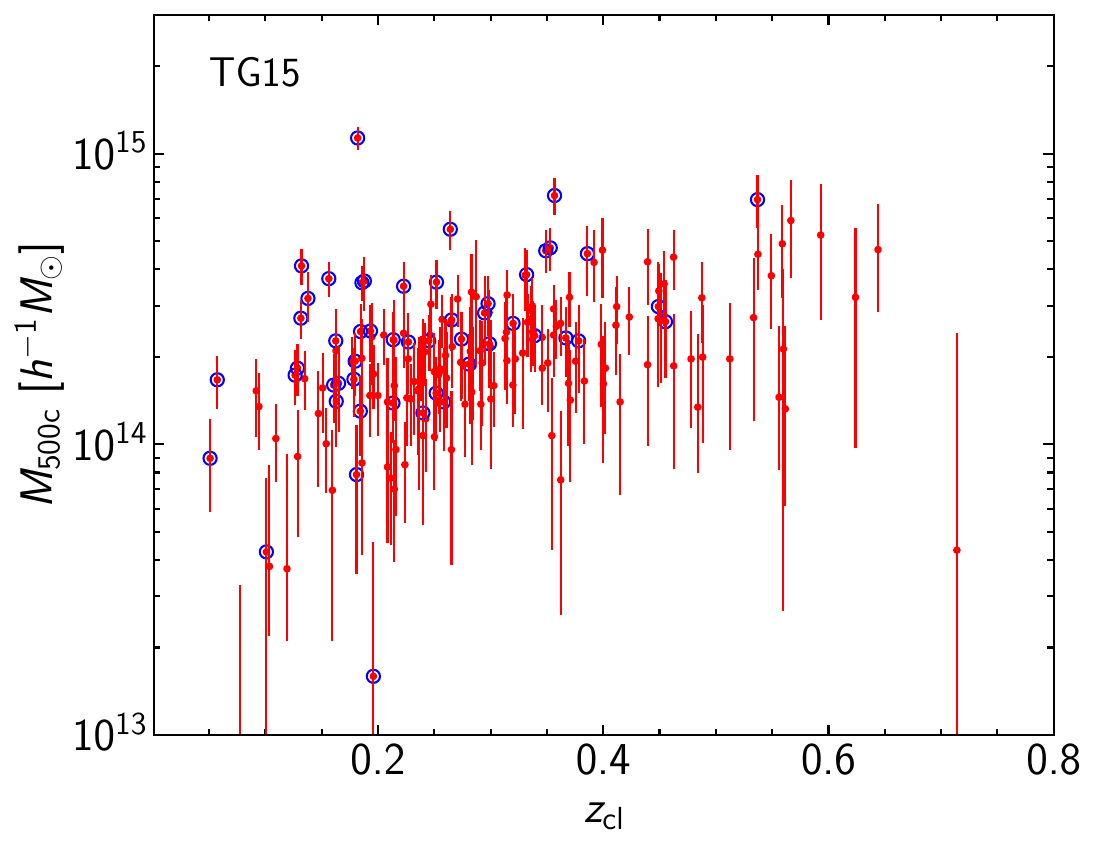}
      \includegraphics[width=0.32\hsize]{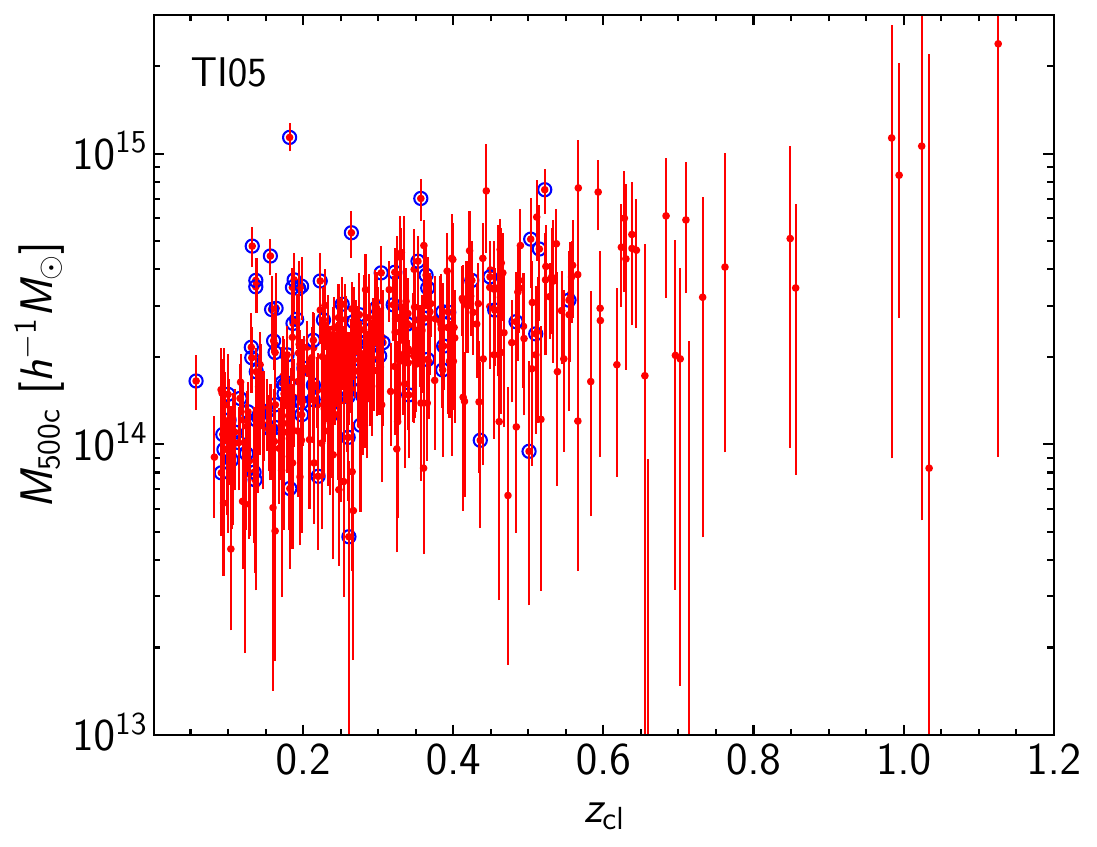}
      \includegraphics[width=0.32\hsize]{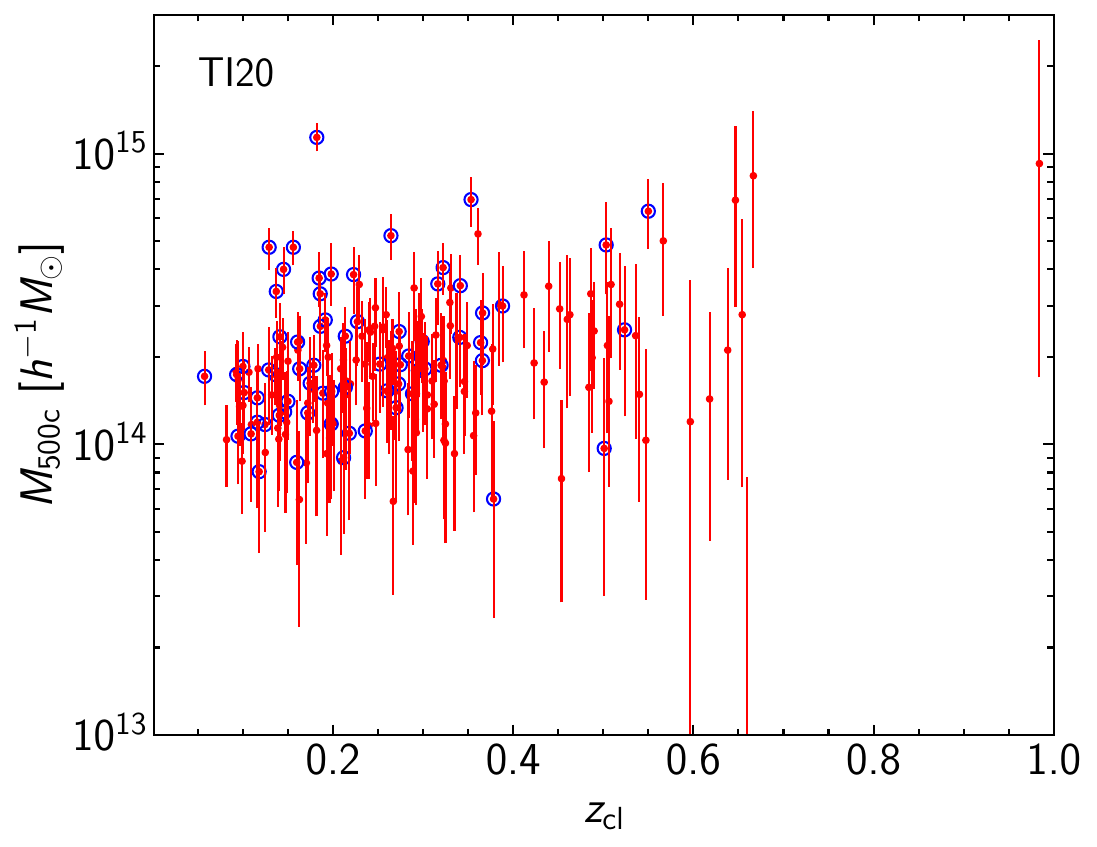}
   \end{center}
  \caption{Weak lensing masses and redshifts of shear-selected
    clusters in the TG15 ({\it left}), TI05 ({\it middle}), and TI20
    ({\it right}) catalogs. Red filled circles with $1\sigma$ errorbars
    show $M_{\rm 500c}$ derived from differential surface density profile
    fitting. Blue open circles indicate clusters that are matched with
    the CODEX catalog.
   }
\label{fig:mass_z}
\end{figure*}

The derived weak lensing masses for the TG15, TI05, and TI20 catalogs
as a function of the cluster redshift are shown in
Figure~\ref{fig:mass_z}.  The fitting results are also given in
Supplementary Tables 1, 2, and 3. The increasing trend of weak
lensing masses with increasing cluster redshift is theoretically
expected for shear-selected clusters \citep[see e.g.,][]{miyazaki18b}.
The massive ($M_{\rm 500c}\sim 10^{15}h^{-1}M_\odot$) cluster at
$z_{\rm cl}\sim 0.18$ is the well-known cluster Abell 1689 that has
one of the largest Einstein radii known \citep[e.g.,][]{oguri09}.
Our weak lensing mass estimation of Abell 1689 is consistent with more
careful lensing mass estimates in the literature \citep[e.g.,][]{umetsu08,umetsu15}.

\begin{figure}
  \begin{center}
      \includegraphics[width=0.95\hsize]{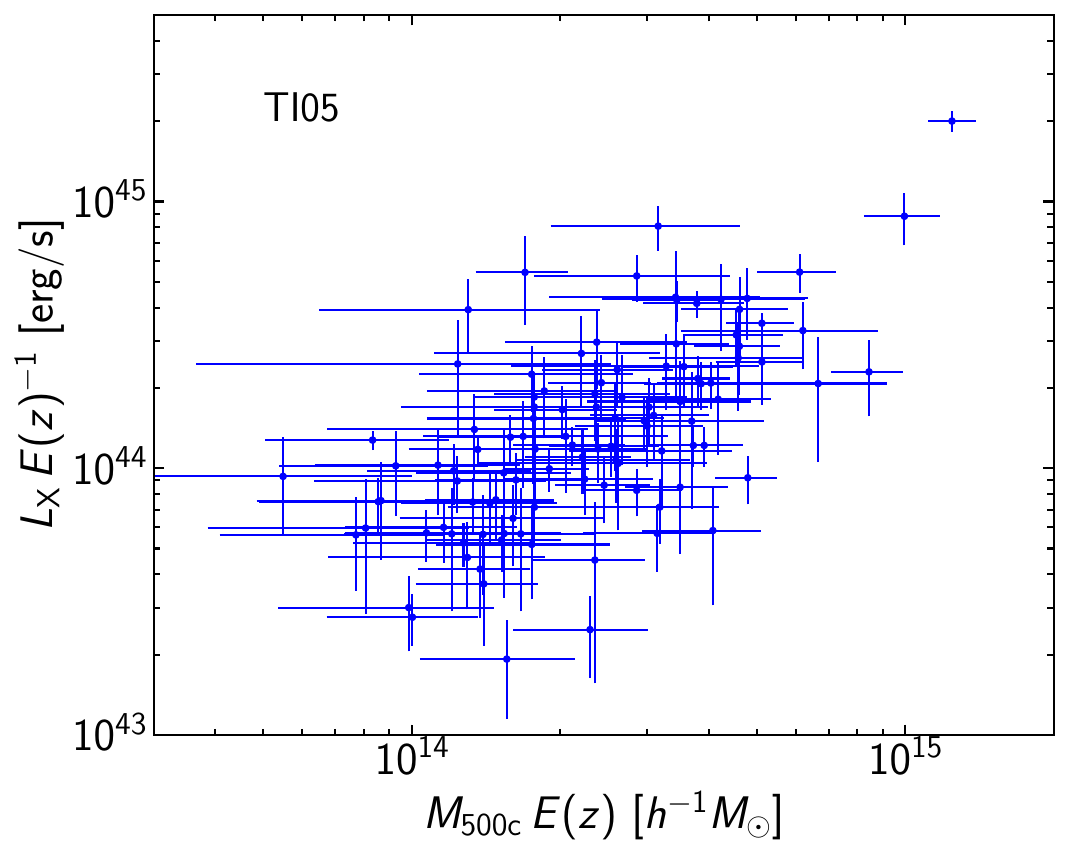}
  \end{center}
  \caption{Weak lensing masses of shear-selected clusters in the TI05
    catalog are compared with X-ray luminosities of CODEX clusters
    that are matched with the shear-selected clusters, with the
    correction of the dimensionless Hubble parameter $E(z)$. The X-ray
    luminosities are measured in the rest-frame $0.1-2.4$~keV
    \citep{finoguenov20}. 
}
\label{fig:mass_lx_codex_ti05}
\end{figure}

In Figure~\ref{fig:mass_z}, we indicate clusters that are matched
with the CODEX catalog. Only less than half of shear-selected clusters
are matched with CODEX clusters (see also Table~\ref{tab:filters}),
which is partly due to the shallow RASS X-ray data that is used to
construct the CODEX catalog \citep{finoguenov20}. As a sanity check,
we compare our weak lensing masses with X-ray luminosities provided by
the CODEX catalog. Figure~\ref{fig:mass_lx_codex_ti05} 
shows the comparison of weak lensing masses with X-ray luminosities
for the TI05 catalog. As expected, we find a good correlation between
the masses and the X-ray luminosities. We note that deriving the
underlying scaling relation requires the correction of selection
biases of both the weak lensing selections in HSC-SSP and the X-ray
selection in RASS. We plan to study X-ray properties of these
shear-selected clusters in more detail using the eROSITA Final
Equatorial Depth  Survey data that are much deeper than the RASS X-ray
data, which will be reported in a separate paper
(Ramos-Ceja et al., in prep.).  

\begin{figure*}
  \begin{center}
      \includegraphics[width=0.32\hsize]{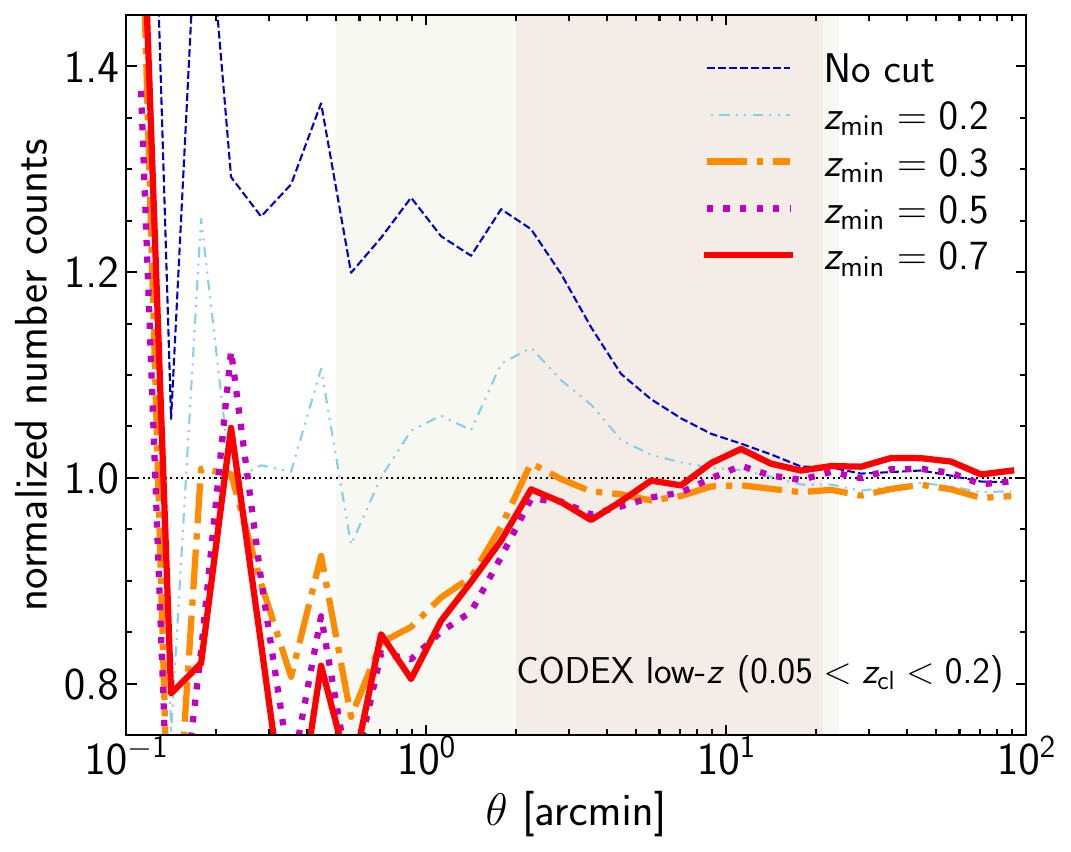}
      \includegraphics[width=0.32\hsize]{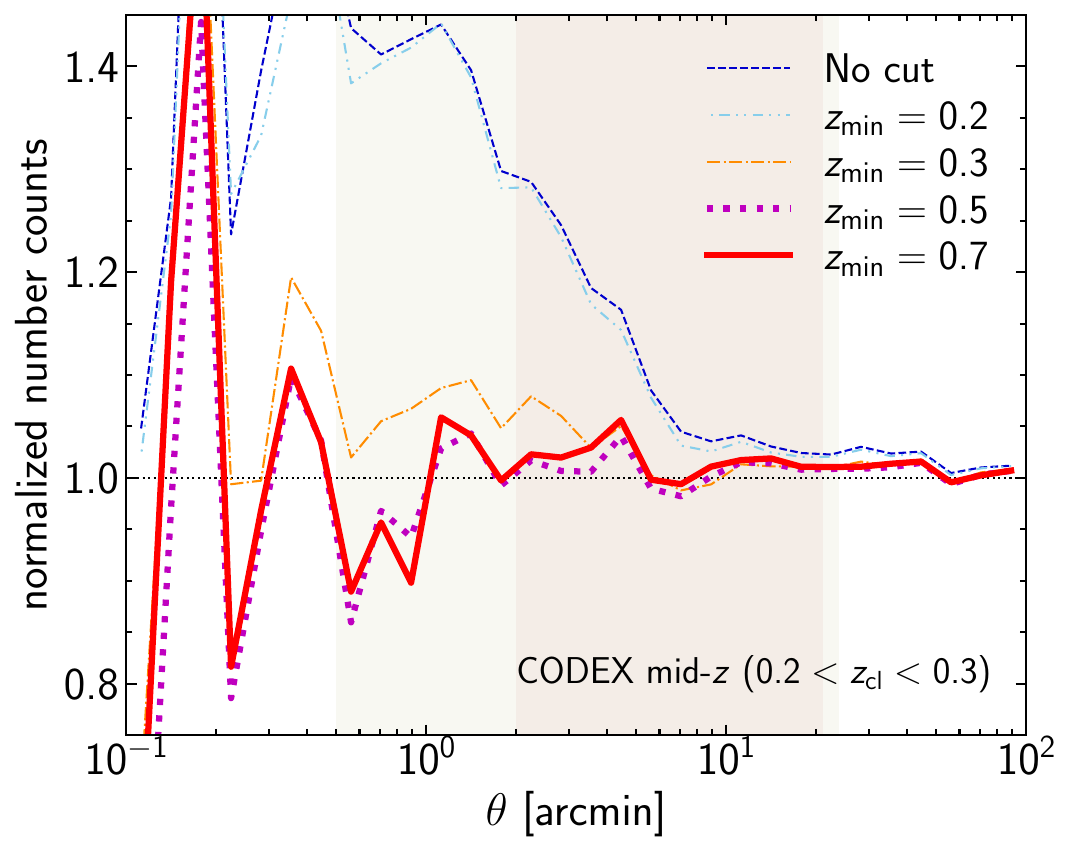}
      \includegraphics[width=0.32\hsize]{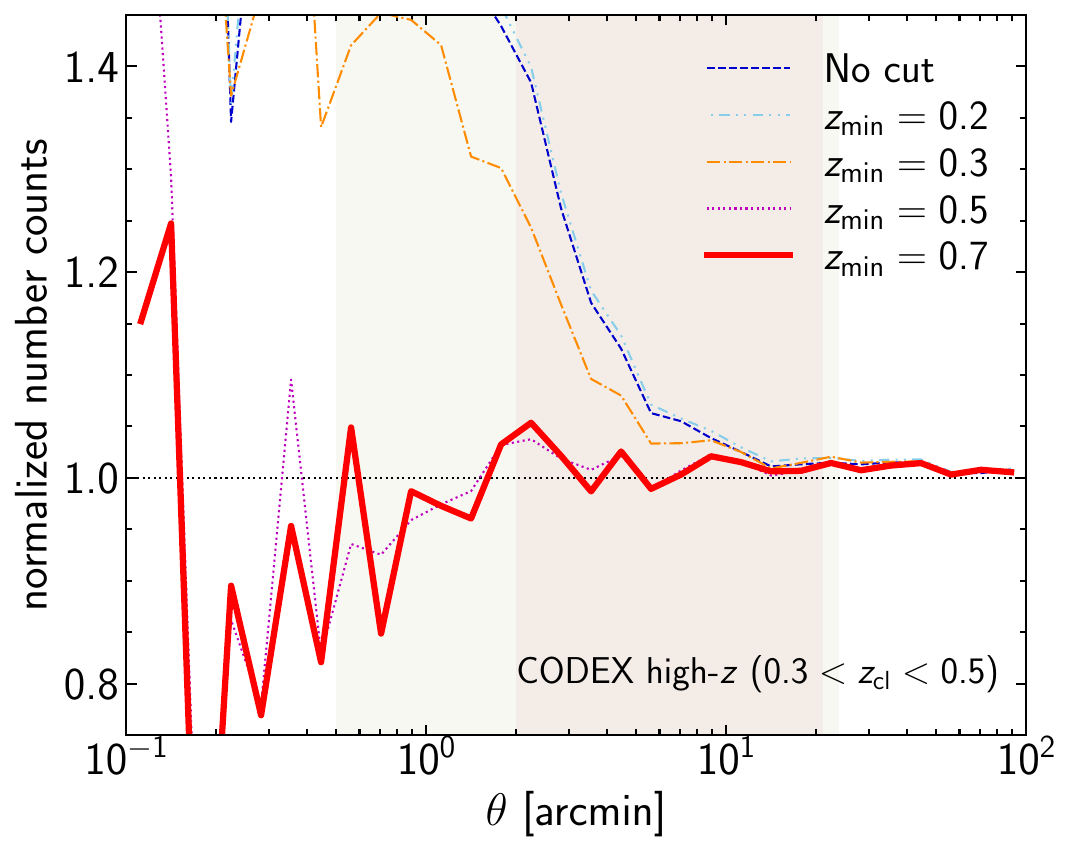}
   \end{center}
  \caption{
    Stacked number density profiles of source galaxies around CODEX
    clusters. Here we show number density profiles relative to those
    from random source catalogs with their total numbers matched to
    total numbers of source galaxy samples such that the
    normalized number density profile should be unity if there is no
    effect of cluster member galaxies. Different lines show
    results for different source galaxy samples summarized in
    Table~\ref{tab:filters}. From left to right panels, stacking is
    conduced around low redshift clusters at
    $0.05<z_{\rm cl}<0.2$ with richness $\lambda>25$, intermediate
    redshift clusters at $0.2<z_{\rm cl}<0.3$ with $\lambda>30$, and
    high redshift clusters at $0.3<z_{\rm cl}<0.4$ with
    $\lambda>40$. Results for background source galaxy samples
    satisfying $z_{\rm min}\geq z_{\rm cl}+0.1$ for all the clusters
    are highlighted by thick lines. The narrow and wide shaded regions
    indicate the non-zero ranges of the kernel function $Q(\theta)$ for
    TI20 and TI05, respectively (see also Figure~\ref{fig:filter}).
  }
\label{fig:numcount}
\end{figure*}

\subsection{Effects of cluster member galaxies}
\label{sec:member}

Cluster member galaxies affect weak lensing measurements mainly in two
ways. One is the enhancement of the number density of source galaxies
due to the contribution of cluster member galaxies, which dilutes weak
lensing signals. The other is the diminishment of the number density
of source galaxies due to the obscuration of small galaxies by cluster
member galaxies. The former effect can be mitigated by using only
source galaxies located behind clusters for weak lensing
measurements. Since these effects are more pronounced near centers of
clusters, choosing the kernel function $Q(\theta)$ that has a smaller
contribution from small $\theta$ can also mitigate the effects of
cluster member galaxies. Here we check number density profiles of our
source galaxy samples around massive clusters to check how our weak
lensing mass maps are affected by cluster member galaxies.

To reduce the statistical noise, we derive stacked number density
profiles of source galaxies for samples of massive clusters. Purely
optically-selected clusters are not ideal for this purpose, because
some of optically-selected clusters exhibit large off-centering up to
$\sim 1$~Mpc, which needs to be taken into account when interpreting
observed stacked number density profiles. Thus we adopt the CODEX
cluster catalog and stack number density profiles around X-ray
centroids of CODEX clusters, because X-ray emission peaks are expected
to be close to halo centers \citep[e.g.,][]{zhang19}.

Figure~\ref{fig:numcount} shows stacked number density profiles for
three CODEX cluster samples with different cluster redshifts.
We also apply the richness cut for each cluster sample in order to select
clusters with many cluster member galaxies, where the richness $\lambda$
for CODEX clusters is measured using the redMaPPer algorithm
\citep{rykoff14}. The richness threshold is
determined so that a sufficient ($\gtrsim 50$) number of clusters are
included in each of the cluster subsamples. We show stacked number
density profiles normalized by those computed with random source
galaxy catalogs in order to highlight effects of cluster member
galaxies. We find a clear signature
of the enhancement of number density profiles without any background
galaxy selection. On the other hand, when source galaxies behind
clusters are selected, we see decrements toward cluster centers due to
the obscuration by cluster member galaxies \citep[or magnification
  effects, see e.g.,][]{chiu20}. We find that decrements are
negligibly small at $\theta> 2'$, suggesting that the TI20 catalog  is
little affected by cluster member galaxy obscurations. In contrast,
the TG15 and TI05 catalogs are more or less affected by cluster member
galaxies in the sense that the observed signal-to-noise ratios may be 
affected by the dilution or obscuration effect due to cluster member
galaxies. Since cluster member galaxies do not contribute to the
signal, the dilution effect enhances the noise and reduce the
signal-to-noise ratio. The obscuration affects both the signal and
noise in a complicated manner.  These effects may be needed to be
taken into account when deriving an accurate selection function of
shear-selected clusters in the TG15 and TI05 catalogs.  

\section{Conclusion}\label{sec:conclusion}

We have constructed shear-selected cluster catalogs by selecting peaks
in weak lensing aperture mass maps covering $\sim 510$~deg$^2$
reconstructed by the HSC-SSP S19A shape catalog. Aperture mass maps
are constructed using the truncated Gaussian filter (TG15) as well as
the truncated isothermal filter with the inner boundary of $0\farcm5$
(TI05) and $2'$ (TI20). For TI05 and TI20, we employ multiple source
galaxy subsamples for which galaxies below redshift $z_{\rm min}$ are
removed to improve the efficiency. With the signal-to-noise ratio
threshold of 4.7, our shear-selected cluster catalogs contain 187, 418,
and 200 clusters for the TG15, TI05, and TI20 set-ups,
respectively. Cross matching with optically-selected cluster catalogs
suggests that the purity of the catalogs is high, more than 95\% for
TG15 and TI20 and more than 91\% for TI05.

These catalogs represent by far the largest catalogs of shear-selected
clusters to date with such high signal-to-noise threshold, and will be
useful for detailed studies of cluster astrophysics and cosmology. In
this paper, we have demonstrated how the shape of the kernel function for
constructing the aperture mass map can be optimized adopting a
flexible functional form of the filter function proposed by
\citet{schneider96}. In particular, we have found that it is possible
to choose the filter function such that it is almost free from effects
of cluster member galaxies yet can select a sufficiently large number
of clusters. Such a clean shear-selected cluster sample will be useful
for obtaining accurate and robust constraints on cosmological
parameters from the cluster abundance, in contrast to
optically-selected clusters for which constraining power appears to be
limited by various systematic effects \citep{abbott20}. We will
explore cosmological constraints with shear-selected clusters in a
forthcoming paper. 

\begin{ack}
We thank T. Hamana and K. Umetsu for useful discussions and comments.
This work was supported in part by the World Premier International
Research Center Initiative (WPI Initiative), MEXT, Japan, and JSPS
KAKENHI Grant Nos. JP18K03693, JP20H00181, JP20H05856.
This work was supported in part by Japan Science and Technology Agency
(JST) CREST JPMHCR1414, and by JST AIP Acceleration Research Grant
No. JP20317829, Japan.

The Hyper Suprime-Cam (HSC) collaboration includes the astronomical communities of Japan and Taiwan, and Princeton University.  The HSC instrumentation and software were developed by the National Astronomical Observatory of Japan (NAOJ), the Kavli Institute for the Physics and Mathematics of the Universe (Kavli IPMU), the University of Tokyo, the High Energy Accelerator Research Organization (KEK), the Academia Sinica Institute for Astronomy and Astrophysics in Taiwan (ASIAA), and Princeton University.  Funding was contributed by the FIRST program from the Japanese Cabinet Office, the Ministry of Education, Culture, Sports, Science and Technology (MEXT), the Japan Society for the Promotion of Science (JSPS), Japan Science and Technology Agency  (JST), the Toray Science  Foundation, NAOJ, Kavli IPMU, KEK, ASIAA, and Princeton University.
 
This paper makes use of software developed for the Large Synoptic Survey Telescope. We thank the LSST Project for making their code available as free software at  http://dm.lsst.org
 
This paper is based on data collected at the Subaru Telescope and retrieved from the HSC data archive system, which is operated by Subaru Telescope and Astronomy Data Center (ADC) at NAOJ. Data analysis was in part carried out with the cooperation of Center for Computational Astrophysics (CfCA), NAOJ.
 
The Pan-STARRS1 Surveys (PS1) and the PS1 public science archive have been made possible through contributions by the Institute for Astronomy, the University of Hawaii, the Pan-STARRS Project Office, the Max Planck Society and its participating institutes, the Max Planck Institute for Astronomy, Heidelberg, and the Max Planck Institute for Extraterrestrial Physics, Garching, The Johns Hopkins University, Durham University, the University of Edinburgh, the Queen’s University Belfast, the Harvard-Smithsonian Center for Astrophysics, the Las Cumbres Observatory Global Telescope Network Incorporated, the National Central University of Taiwan, the Space Telescope Science Institute, the National Aeronautics and Space Administration under grant No. NNX08AR22G issued through the Planetary Science Division of the NASA Science Mission Directorate, the National Science Foundation grant No. AST-1238877, the University of Maryland, Eotvos Lorand University (ELTE), the Los Alamos National Laboratory, and the Gordon and Betty Moore Foundation.
 \end{ack}

\bibliographystyle{apj}
\bibliography{refs}

\appendix
\section{Optimization of the truncated isothermal filter}\label{app:ti_filter}

We use signal-to-noise ratios of mass map peaks computed assuming the
NFW profile \citep{navarro97} to optimize parameters of the truncated
isothermal filter presented in subsection~\ref{sec:filter_ti}.
For each set of parameters, we compute the aperture mass $M_{\rm ap, NFW}$
at the center of a halo using equation~(\ref{eq:m_ap_1}) with the
convergence assuming an NFW profile \citep[e.g.,][]{bartelmann96}. For
each subsample of source galaxies given in Table~\ref{tab:filters}, we
fully take account of the source redshift distribution by adopting the
stacked PDF of the photometric redshift PDFs of all the source
galaxies. We also adopt the fitting form of the mass-concentration
relation presented by \citet{diemer15} and \citet{diemer19}. 
The signal-to-noise ratio is then computed as
\begin{equation}
  \nu_{\rm NFW}=\frac{M_{\rm ap, NFW}}{\sigma_{\rm shape}},
  \label{eq:nu_nfw_def}
\end{equation}
where $\sigma_{\rm shape}$ describes the shape noise of the filtered
convergence field that is computed as
\begin{equation}
\sigma_{\rm shape}=\sigma_e\sqrt{\frac{\pi\int d\theta\,\theta\,Q^2(\theta)}{n_{\rm gal}}},
\end{equation}
with $\sigma_e$ and $n_{\rm gal}$ being the root-mean-square of the
ellipticity and the number density of source galaxies, respectively.
In this calculation we simply assume $\sigma_e=0.4$ and
$n_{\rm gal}=22$~arcmin$^{-2}$ before any source galaxy selection so
that the resulting noise of the mass map roughly coincides with that
from the real data. For each source galaxy selection we reduce
$n_{\rm gal}$ according to the weighted sum of the number of galaxies
after the source galaxy selection. 

While it is customary to define a sample of shear-selected clusters by
applying a threshold to the signal-to-noise ratio that is defined in a
manner similar to equation~(\ref{eq:nu_nfw_def}) where only the shape
noise is considered, it is known that the accumulated density
fluctuations along the line-of-sight (i.e., cosmic shear) also
contribute to the noise.
\begin{equation}
\sigma_{\rm LSS}=\sqrt{\int \frac{\ell d\ell}{2\pi}|\hat{U}(\ell)|^2C_\ell},
\end{equation}
where $\hat{U}(\ell)$ is the Fourier counterpart of the filter
$U(\theta)$ and the cosmic shear power spectrum $C_\ell$ is related to
the nonlinear matter power spectrum $P_{\rm m}(k;z)$ as 
\begin{equation}
C_\ell =\int d\chi
\frac{\left[W^\kappa(\chi)\right]^2}{\chi^2}P_{\rm m}(k=\ell/\chi; z),
\end{equation}
\begin{equation}
W^\kappa(\chi)=\int_z^\infty dz_{\rm s}\frac{4\pi G}{c^2}\frac{(\chi_{\rm
    s}-\chi)\chi}{\chi_{\rm s}(1+z)^2}\bar{\rho}_{\rm m}p(z_{\rm s}),
\end{equation}
where $\chi$ and $\chi_{\rm s}$ are the comoving radial distances
corresponding to redshift $z$ and $z_{\rm s}$, respectively, and
$p(z_{\rm s})$ denotes the redshift distribution of source galaxies.
We use the revised {\tt halofit} model of \citet{takahashi12} to
compute $P_{\rm m}(k;z)$. The signal-to-noise ratio including the
large-scale structure noise is simply calculated as
\begin{equation}
  \nu_{\rm NFW, wLSS}=\frac{M_{\rm ap, NFW}}{\sqrt{\sigma_{\rm
        shape}^2+\sigma^2_{\rm LSS}}}.
  \label{eq:nu_nfw_lss_def}
\end{equation}

We optimize parameters of the truncated isothermal filter as
follows. For each source galaxy selection listed in
Table~\ref{tab:filters}, we consider NFW halos located at
$z=z_{\rm min}-0.1$ with varying halo mass as representative halos
detected in mass maps with the source galaxy selection characterized
by $z_{\rm min}$.
For each mass of the NFW halo, we vary $\nu_1$, $\nu_2$, and
$\theta_R$ to search for the  optimal set of parameters that maximize
$\nu_{\rm NFW, wLSS}$ given by equation~(\ref{eq:nu_nfw_lss_def}).
Since the combination of $\nu_1\theta_R$ determines the inner boundary
of the filter (see subsection~\ref{sec:filter_ti}), in this paper we
consider two cases, $\nu_1\theta_R=0\farcm 5$ and $2'$. The former is
chosen such that the tangential shear at $\theta\sim 1'-1\farcm5$,
where the contribution to the signal is large, is included. The latter
removes the significant fraction of the inner part of the profile from
the calculation, and hence is much less affected by various systematic
effects as discussed in subsection~\ref{sec:filter_ti} (see also
subsection~\ref{sec:member}). First we vary all the three parameters
with the constraint on $\nu_1\theta_R$ to find that
$\nu_{\rm NFW, wLSS}$ is generally maximized for 
$\nu_2\sim 0.3-0.4$. We thus fix $\nu_2=0.36$ throughout the paper and
derive the optimal choice of $\nu_1$ and $\theta_R$, as well as
$\nu_{\rm NFW}$ and $\nu_{\rm NFW, wLSS}$, as a function of the halo 
mass. We adopt values of $\nu_1$ and $\theta_R$ for the halo mass that
yields $\nu_{\rm NFW}\sim 5$, roughly corresponding to threshold of
constructing shear-selected clusters used in the
literature. Parameters determined by this procedure for each
$\nu_1\theta_R$ and the source galaxy selection are presented in
Table~\ref{tab:filters}.  

\section{Supporting Information}
Supplementary Tables 1, 2, and 3 are available online.

\end{document}